\documentstyle[12pt]{article}
\newcommand{\beq}{\begin{equation}}
\newcommand{\eeq}{\end{equation}}
\newcommand{\bea}{\begin{eqnarray}}
\newcommand{\eea}{\end{eqnarray}}
\def\bar{\begin{array}}
\def\ear{\end{array}}

\newcommand{\G}{{\cal G}}
\def\le#1{\label{eq:#1}}
\def\re#1{\ref{eq:#1}}

\begin{document}

\noindent
\centerline{\bf Dispersive effects in neutron matter superfluidity}

\vskip 0.5 cm

\noindent
\centerline{
M. Baldo$\dag$ and A. Grasso$\ddag$}

\begin{small}

\vskip 0.5 cm
\noindent
\centerline{ $\dag$ INFN, Sez. Catania, 
Corso Italia 57 - 95129 Catania, Italy}

\noindent
\centerline{ $\ddag$ Dipartimento di Fisica, Universita' di Catania,
Corso Italia 57, 95129 Catania, Italy}

\vskip 1 cm

\noindent
{\bf Abstract.} The explicit energy dependence of the single particle 
self-energy (dispersive effects),
due to short range correlations, is included in the treatment of neutron
matter superfluidity. The method can be applied in general to strong 
interacting fermion systems, and it is expected to be valid whenever 
the pairing gap is substantially smaller than the Fermi kinetic energy. 
The results for neutron matter show that dispersive effects are strong 
in the density region near the gap closure.

\vskip 0.6 cm

\noindent{\bf PACS}: 21.65+f , 26.60+c , 97.60Jd        
\vskip 0.8 cm

\end{small}

\vskip 0.8 cm

\noindent
{\bf 1.~Introduction}
\vskip 0.3 cm
 
\noindent

Neutron and nuclear matter superfluidity is one of the main issue in 
the physics of
neutron stars. Superfluidity is expected to play a major role in some
of the most striking phenomena occurring in neutron stars, like 
glitches and post-glitches transients \cite{shap}, vortex pinning \cite{pines},
neutron star cooling, and maybe strong magnetic field penetration \cite{baiko}. 
However, since the observational data are only indirectly related to 
superfluidity, or need explicit models for their interpretation,
a firm theoretical prediction of the superfluidity
strength, based on microscopic {\it ab initio} calculations, appears 
highly required. Unfortunately, neutron and nuclear matter are strongly 
correlated systems, where short range correlations dominate the overall 
interaction energy, even at densities well below the saturation value.
The superfluidity problem turns out, therefore, to be a complex many-body
problem, where the delicate balance between short range interactions
and the long range pairing  correlations needs an  accurate
treatment. This problem was firstly considered in the works of
ref. \cite{clark1}, within the variational Jastrow method. 
The medium effect on the effective pairing interaction was investigated
in refs. \cite{clark2,hans}. The correlation effects were treated in ref.
\cite{wambach} within a generalization of the Babu-Brown approach
\cite{bab} to the effective nucleon-nucleon (NN) interaction and in the weak
coupling limit. In general, these microscopic approaches seem to indicate 
 a reduction of the pairing gap due to the medium,
with respect to BCS approximation with the bare interaction.
The use \cite{baldo} in the BCS scheme of realistic bare nucleon-nucleon 
interactions, which reproduce the experimental phase shifts, 
can be a good starting point for a more sophisticated many-body
treatments, and the connection between the pairing gap
value and the phase shifts has been elucidated, in general, 
in ref. \cite{martino}.\par
In all these microscopic approaches the
single particle spectrum is usually considered within the effective 
mass approximation, or taken from normal state calculations.
Dispersive effects, due to the energy dependence of the single particle
self-energy, are usually neglected, or considered
only in the weak coupling limit, and no first principle scheme to include 
them in the general gap equation has been proposed. Only recently \cite{bozek}
a self-consistent scheme has been developed, where the short range 
correlations and the pairing problem are treated on the same footing.
The method is numerically complex, and it has been
solved only for a schematic interaction \cite{bozek}. 
It would be desirable to understand in simple terms the effect of
the self-energy dispersion on the pairing strength, also in the case
of strong coupling as in neutron matter. In this letter 
we present a general scheme for including dispersive effects in the gap
equation, which is simple and accurate, provided the gap
is substantially smaller than the Fermi kinetic energy. We then apply 
the method to neutron matter superfluidity with realistic interaction, 
and we show numerically that the size of the effect can be 
large in the vicinity of the gap closure. 
\vskip 0.5 cm
\noindent
{\bf 2.~Including  self-energy in the gap equation}
\vskip 0.3 cm
\noindent
The general many-body theory of pairing in fermion systems has been
formulated just after the BCS \cite{BCS} solution has been introduced, 
and it can be found in standard textbooks {\cite{noz,abri}}.
The main framework is the Green' s function (GF) formalism, which
generalizes the Gorkov' s method beyond the BCS approximation. The single 
particle Green's function $\G$ has a 2x2 matrix structure, with normal 
diagonal components $F_1$ and abnormal off-diagonal components $F_2$,
\beq  
 \G (k,\omega) = \left( { \bar{cc} {\!\!\!\!\!\!\!\! F_1(k,\omega)} \ 
                            \, \ {F_2(k,\omega)} \\
                                { F_2(k,\omega)} \ -{F_1(k,-\omega)}  \ear }  
\right)
 \, , \, 
 \G^{-1} (k,\omega)\, =\, \left( {\bar{cc} 
 {\!\!\!\!\!\!\!\! \tilde{\epsilon}_k - \omega +
 M(k,\omega)} \ \,\, {\Delta (k,\omega)} \\
 {\Delta (k,\omega)} \ \,\,
 {-(\tilde{\epsilon}_k + \omega + M(k,-\omega))} 
  \ear    } \right)
\le{green}
\eeq
\noindent
In the expression for the inverse Green's function $\G^{-1}$, we
have introduced  the quantity $\tilde{\epsilon}_k = \hbar^2 k^2/2m - \mu$ 
as the single 
particle kinetic energy, with respect to the chemical potential $\mu$, 
the diagonal single particle self-energy $M(k,\omega)$ and 
the momentum and energy dependent gap function $\Delta (k,\omega)$.
Here we are assuming s-wave singlet pairing, and therefore we omit
spin indices. They simply express the coupling between the time-reversal 
states $(k,\uparrow)$ and $(-k,\downarrow)$.
Both $M(k,\omega)$ and $\Delta (k,\omega)$ can be expanded in terms
of the NN interaction and the full GF itself, which in general entails
a self-consistent procedure. The gap function $\Delta (k,\omega)$,
however, is solution of the homogeneous generalized Bethe-Salpeter equation
\cite{noz}, and as such it satisfies the generalized gap equation
\beq  
 \Delta (k,\omega) \, =\, \sum_{k'}\int d\omega' 
 {I(k\omega,k'\omega') \, \Delta (k',\omega')
\over (\tilde{\epsilon}_{k'} - \omega' + M(k',\omega')) 
         (\tilde{\epsilon}_{k'} + \omega' + M(k,-\omega')) + 
         \Delta (k',\omega')^2}
\le{gap}
\eeq
\noindent
where $I(k\omega,k'\omega')$ is the irreducible NN interaction at
zero total energy and momentum. If one takes 
the bare NN interaction for the interaction $I$,
and the Hartree-Fock approximation for the diagonal self-energy $M(k,\omega)$,
the standard BCS approximation is
recovered. It has to be noticed that the energy dependence of the 
gap function $\Delta (k,\omega)$ originates only from the energy dependence
of the irreducible interaction $I$. In fact, if the interaction is taken
as energy-independent, the gap function is also energy-independent,
despite the possible energy dependence of the self-energy $M(k,\omega)$.
Since we are looking for an estimate of the dispersive effects,
we indeed assume the irreducible interaction as energy independent,
while we keep the full energy dependence of the self-energies.
Then, the energy integration appearing in the gap equation (\re{gap})
can be performed with a good accuracy in the limit of small self-energy 
imaginary part, since than the main contribution is expected
to come from the poles close to the real axis. 
The denominator
is an even function of the energy $\omega$ (we remind again that single
particle energies
are measured with respect to $\mu$), and, therefore, the kernel presents
two poles, symmetrical with respect to the origin in the
complex $\omega$-plane. This is a feature typical of the 
superconducting phase. Formally, the pole energies $\pm E_k$ are 
the solutions of the implicit equation
\beq
\pm E_k = {1\over 2} (M(k,\pm E_k) - M(k,\mp E_k)) \pm
 \sqrt{ \left[ \tilde{\epsilon}_k + {1\over 2} (M(k,-E_k) + M(k,E_k))\right]^2
 + \Delta(k)^2 }
\le{impl}
\eeq
\noindent
If the energy dependence of $M(k,\omega)$ is neglected, than 
Eq. (\re{impl}) reduces to the usual square root expression for the
quasi-particle excitation energy of the BCS approximation. On the
other hand, 
in the non-superconducting limit $\Delta \rightarrow 0$, 
and neglecting the imaginary part of $M(k,\omega)$, one can verify
that Eq. (\re{impl}) reduces to the usual self-consistent
equation, e.g. Brueckner \cite{book}, for the single particle energy $e_k$
\beq
     e_k = \tilde{\epsilon}_k + M(k,e_k)
\le{bru}
\eeq
\noindent
Equation (\re{bru}) is valid whenever $\Delta$ is negligible, 
in particular for momenta far away from the Fermi surface,
since then $|\Delta_k| \ll |\tilde{\epsilon}_k|$.
We now make use of the assumption of a gap $\Delta$ smaller 
than the normal self-energy, which is mainly determined
by short range correlations. The size of the normal self-energy
is indeed of the same order of the Fermi kinetic energy $E_F$.
In this case, Eq. (\re{bru}) will be valid to order $\Delta(k_F)/ E_F$,
and therefore on the right hand side of Eq. (\re{impl}) we can 
replace $E_k$ with $e_k$, solution of Eq. (\re{bru}), to get
\beq 
E_k \approx {1\over 2} (M(k,|e_k|) - M(k,-|e_k|)) +  
 \sqrt{ \left[ \tilde{\epsilon}_k + 
 {1\over 2} (M(k,-e_k) + M(k,e_k))\right]^2
 + \Delta(k)^2 }
\le{res1}
\eeq
\noindent
where the self-energy is now calculated in the normal phase.
The procedure is justified,
provided $M(k,\omega)$ is a smooth function of $\omega$. 
It will be further discussed below.
\par
The residue $R$ of the kernel at each one of the pole can be easily calculated
\bea
 R = \left[ 1 - 
 {1\over 2}( (1 - \Theta_k) a_k + (1 + \Theta_k) b_k )\right]^{-1} \\
 \Theta_k = {\tilde{\epsilon}_k + {1\over 2}(M(k,-E_k) + M(k,E_k))
        \over E_k } \\
 a_k \, = \, \left({\partial M\over \partial \omega}\right)_{\omega = E_k} 
 \ \ \ \ \ ; \ \ \ \ \
 b_k \, = \, \left({\partial M\over \partial \omega}\right)_{\omega = -E_k}
\eea
\noindent
In the limit $\Delta \rightarrow 0$, Eq. (6) is the usual
expression for the quasi-particle strength $Z_k$, provided the
momentum $k$ is close enough to $k_F$. 
The corrections to the normal phase value of $1 - R^{-1}$ are of the order 
$\Delta(k_F)/ E_F $, and therefore the residue $R$ can  be identified 
 with $Z_k$, at least in the vicinity of the Fermi momentum. 
Far away from the Fermi
momentum, the residue $R$ has still the expression of Eq. (6)
in the limit of small imaginary part, despite the quasi-particle
concept becomes less meaningful, since its width can be large (but
it can be still much smaller than the real part of the energy).
In this case the procedure is just an accurate method of calculating
the energy integral (pole approximation). For simplicity, the residue 
will be denoted by $Z_k$ in all cases. 
 Within these approximations, the generalized gap equation
(\re{gap}) reads
\beq
\Delta (k) \, = \, -\sum_{k'} I(k,k') Z_{k'} {\Delta_{k'}\over 
 2 \sqrt{ \left[ \tilde{\epsilon}_k +
 {1\over 2} (M(k,-e_k) + M(k,e_k))\right]^2
 + \Delta(k)^2 } }
\le{gapa}
\eeq
\noindent
Eqs. (\re{res1}), (6), (\re{gapa}) contain the main result of the paper.
It has to noticed that in the generalized gap equation (\re{gapa}) 
the square root in the denominator does not coincide with the
quasi-particle energy of Eq. (\re{res1}), in contrast with the usual BCS
approximation. This feature is general and it is not bound to the 
approximation of Eq. (\re{res1}). The approximation
of Eq. (\re{res1}) to the expression inside the square root is valid 
up to corrections of order $\Delta/E_F$. They can be absorbed in large part  
by a small shift of the chemical potential $\mu$.
\par
 Before going to the application of the formalism to
neutron matter, let us discuss Eq. (\re{gapa}) in the extreme weak 
coupling limit, where one assumes that the main contribution to the
momentum integral is concentrated around the Fermi surface. 
In this limit, following the standard procedure
of expanding the integrand of Eq. (\re{gapa}) around $k_F$, 
one gets \cite{fetter}
\beq
\Delta_F \, = \, 8 {E_F\over m_F} \exp (-{1\over m_F Z_F\pi^2 n_0 I(k_F)})
\le{weak}
\eeq
\noindent
where $n_0$ is the density of state for the free Fermi gas and
$m_F$ the so called $k-mass$ (in units of the bare mass) \cite{neg}.
The 
interaction $I(k_F)$ is the diagonal matrix element of the NN potential
in the considered channel ( $^1S_0$ for neutron matter), in the
plane wave representation. The self-energy effects are,
therefore, contained mainly in the factor $m_F Z_F$, which can be written also
as $m^* Z_F^2$, since the full effective mass $m^* = m_F/Z_F$ \cite{neg}.
This is the standard result for the weak coupling limit \cite{migdal}.
Eq. (\re{gapa}) generalizes the treatment to the case where the 
contribution from momenta far from the Fermi momentum is relevant.
The appearance of the $k-mass$ is a peculiar feature of the
pairing phenomenon, and it is a direct consequence of the coupling
between time-reversal states. In Eq. (\re{gapa})
the combination  $M(k,-\omega)+M(k,\omega)$ gives rise to the
{\it combined} density of state of the pair $\{ (k,\omega) ; (-k,-\omega)\}$,
which is mainly determined by the $k-mass$.\par
The weak coupling limit is not valid in general for neutron
or nuclear matter \cite{book}, if one starts from the bare
NN interaction. This can be seen directly from
the observation that often the gap equation has a well defined solution
even when the interaction matrix element $I(k_F)$ is positive. This 
is due to the dominant role of the off-diagonal matrix elements $I(k,k')$. 
Therefore, one must solve the more
general equation (\re{gapa}) in this case. The above considerations
are, anyhow, still valid.
\newpage
\vskip 0.5 cm
\noindent
{\bf 3.~Application to neutron matter superfluidity}
\vskip 0.3 cm
\noindent
In order to estimate the dispersive effects on the superfluid gap
of neutron matter, we have solved Eq. (\re{gapa}), with the bare 
Argonne v$_{14}$ potential as the pairing interaction $I(k,k')$ and
with the self-energy calculated in the Brueckner approximation at the
lowest order, see Fig. 1a, with the same interaction. The higher
order contribution of Fig. 1b turns out to be indeed negligible in
the relevant density range.  
In the superfluid phase, in principle,
the diagonal self-energy $M(k,\omega)$ differs from the
self-energy in the normal phase. The main contribution 
not present in the normal phase originates from the coupling of the
single particle motion with the superfluid collective modes. The latter
correspond mainly to the center of mass motion of the Cooper
pairs and their possible ``vibrations" \cite{old,wolfle}. 
The branch starting at zero energy, in the long wave-length limit,
is the branch of the Goldstone boson \cite{old}, corresponding to the gauge
invariance symmetry breaking at the superfluid phase transition.
This contribution to the diagonal single particle self-energy
is expected to be at most of the order of the superfluid condensation 
energy per particle, and therefore negligible with respect to the
typical short range correlation energy, as calculated  e.g. in
Brueckner theory, at least to the extent that $\Delta/E_F \ll 1$.
For the same reason, 
the deviation of the occupation number from the free gas
value and the presence of a forbidden energy region,
of order $\Delta$, around the Fermi energy, typical of the pairing phenomenon,
seem to play no relevant role in determining the size of the self-energy.
It appears, therefore, justified to adopt for $M(k,\omega)$ its
normal phase value.
\par
The choice of the bare interaction for $I(k,k')$ is suggested
by the observation that no ladder summation should be included in the
irreducible interaction kernel $I(k,k')$ \cite{migdal,book}. Of course,
other terms, like polarization diagrams, should be included \cite{hans},
but here we want simply to single out the dispersive effects, and therefore
it appears meaningful to compare the results obtained with and without 
self-energy, within the same scheme of approximation. \par
In Fig. 2 is reported, for $k_F = 0.9 fm^{-1}$, the imaginary part 
of the neutron self-energy $M_I(k,e(k))$ at the quasi-particle pole,
as a function of the momentum $k$, together with the real 
part of the quasi-particle energy $e(k)$ (calculated with respect to $\mu$). 
One can see that indeed the imaginary part is small with respect to the
real part. The situation is completely similar for the other densities.
The residue $Z_k$ of Eq. (6), which appears in 
Eq. (\re{gapa}) for the gap function, is reported in Fig. 3 for 
three densities. According to Migdal-Luttinger theorem \cite{migdal}, 
the value $Z_F$ of $Z_k$ at $k = k_F$ is the discontinuity of the momentum
distribution at the Fermi momentum (in the normal phase). One must have, 
therefore, $0 < Z_F < 1$.  
One can see, however,  that $Z_k$  exceeds 1 slightly in some
interval well above $k_F$. This is not surprising, since for large $\omega$
values, at fixed $k$, the real part of the self-energy is an increasing 
function of $\omega$, and it
approaches asymptotically an energy-independent value \cite{mahaux}
(and therefore $Z_k \rightarrow 1$). As already mentioned, for large momentum
the pole approximation is just an accurate method of calculating the 
relevant energy integral. The position of the pole is, of course,
not exactly on the real axis, but numerical estimate of the second derivative
of the self-energy shows that to calculate $Z_k$ on the real axis
is an extremely good approximation. The factor $Z_k$ is also related
to the single particle occupation number $n(k)$.
In fact, the contour integral, closed  in the lower complex plane,
of the single particle GF equals $1 - n(k)$, $k > k_F$. 
Besides the pole, this integral receives contribution from 
the regular (non polar) part of the single particle GF
 \cite{migdal}. If $Z(k) > 1$, this means simply that
the regular part leads to a slightly negative contribution.
We have checked, indeed, that the single particle spectral function,
calculated with the same self-energy (including of course the imaginary part),
satisfies the sum rules and gives well defined occupation numbers $n(k)$.
A full account of the calculations will be reported elsewhere.
Anyhow, this small deviation from 1 does not affect at all 
the final results, and one can take $Z_k = 1$ in this momentum region. 
\par
Finally, in Fig. 4  is reported the pairing gap
at the Fermi energy as a function of density for three different
cases : i) without self-energy \cite{baldo}, ii) with 
the $Z_k$ factor in the numerator of the gap equation (\re{gapa}), 
iii) with both the self-energy in the denominator
and the $Z_k$ factor. The reduction of the pairing gap
is substantial at the highest densities, near the gap closure. 
\vskip 0.5 cm
\noindent
{\bf 3.~Discussion and conclusion}
\vskip 0.3 cm
\noindent
 We have developed a method to include the single particle self-energy in
 the gap equation, and in particular dispersive effects, beyond the
 usual BCS approximation. The method rely on the assumption of a small
 gap with respect to the Fermi kinetic energy and  strong short range
 correlations, typical of the neutron matter in the inner crust of neutron 
 stars. The results indicate that dispersive effects can strongly
 reduce the gap value near its closure. The effect is due both to the
 quasi-particle strength $Z_k$ and to the $k-mass$,
 which enter in the generalized gap
 equation (\re{gapa}). This result appears in line with the work
 of ref. \cite{bozek}, where the self-consistent treatment of 
 pairing and short range correlations seems indeed to reduce strongly the
 gap value mainly because of these two factors \cite{bozek,private}.
 Of course, before drawing any conclusion on the pairing strength
 in neutron matter, one should include, along a consistent scheme,
 the correlation effects on the irreducible interaction $I(k,k')$.
  Work in this direction is in progress. 
  \par
 The extension of the method to symmetric nuclear matter,
 possibly relevant for pairing in nuclei, appears
 problematic. In that case the approximation of a small imaginary part
 looks less justified away from the
 Fermi momentum \cite{mahaux}. Furthermore, the neutron-proton
 pairing in the $^3S_1 - ^3D_1$ channel is too strong, i.e. $\Delta \sim E_F$,
 to be treatable in the proposed approximate method. In both cases
 a self-consistent procedure, involving both the self-energy and the 
 effective interaction, seems to be the only viable method. 
\newpage
\noindent
{\bf Figure captions}
\bigskip
\par\noindent
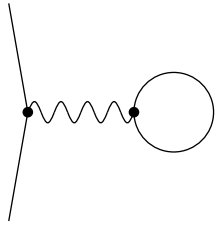
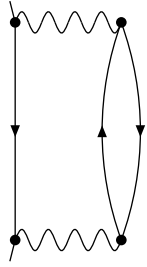
Fig. 1.- One (a) and two (b) hole-line diagrams contributing to the
nucleon self-energy in the Bethe-Brueckner-Goldstone expansion.
The wavy lines indicate Brueckner G-matrices.
\par\noindent
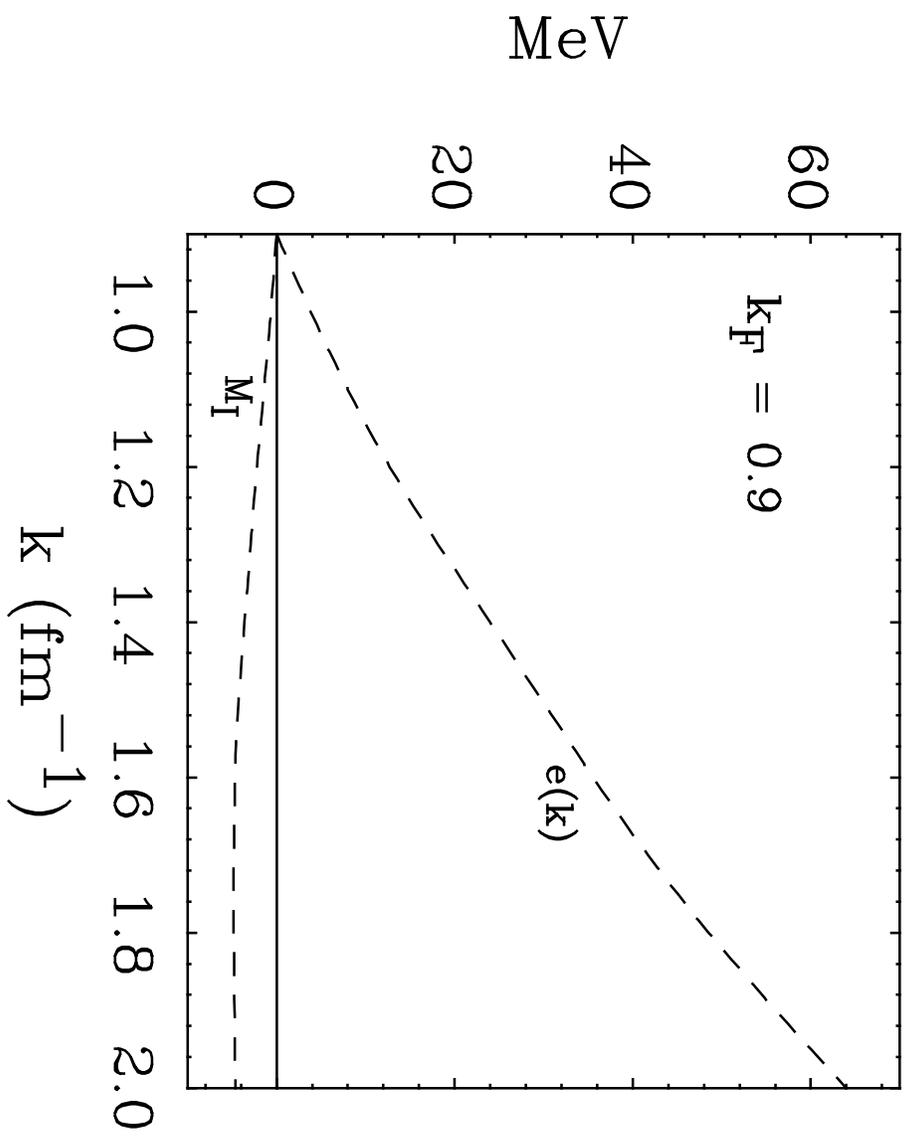
Fig. 2.- Real part e(k) and imaginary part $M_I$ of the quasi-particle
pole, as a function of the momentum $k$ at the Fermi momentum $k_F = 0.9
fm^{-1}$.
\par\noindent
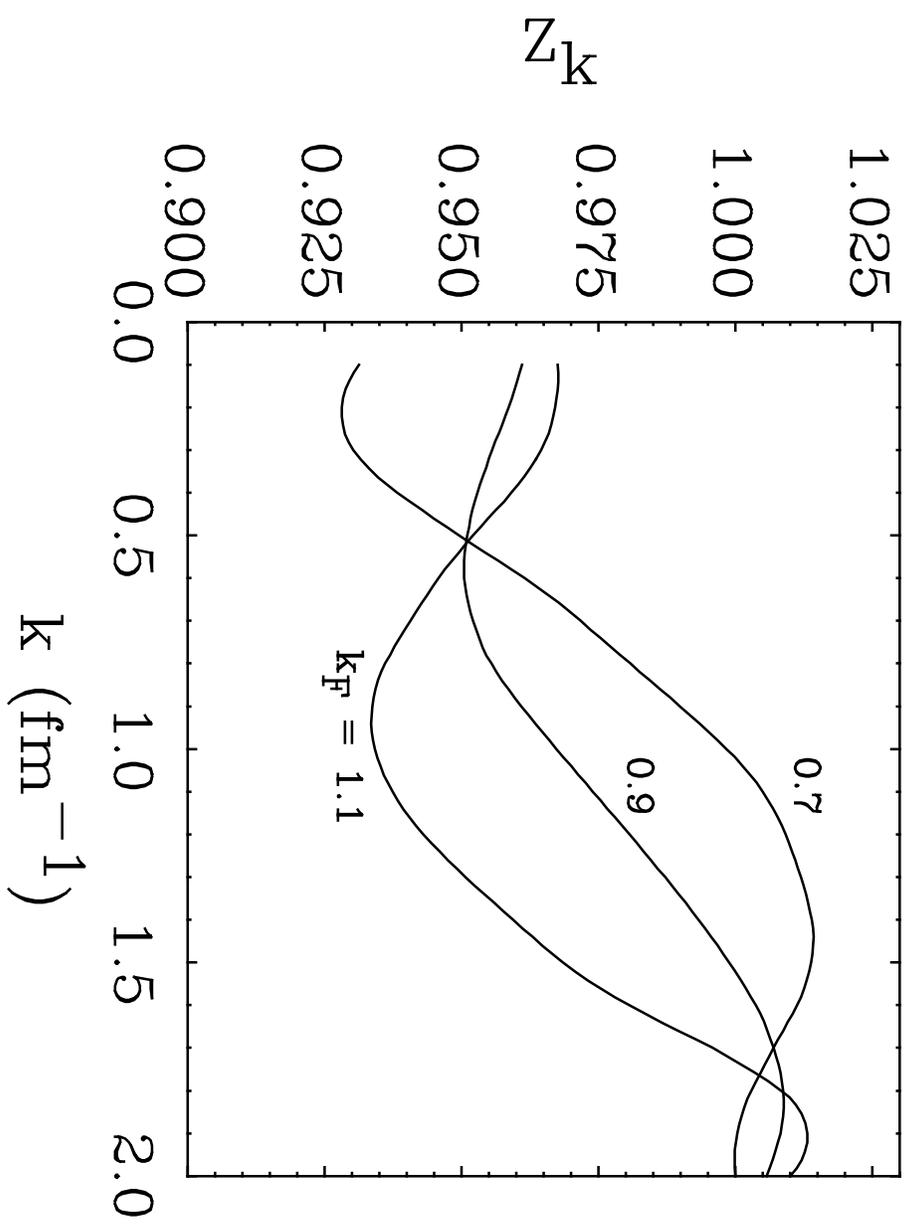
Fig. 3.- The residue $Z_k$ at the quasi-particle pole as a function
of the momentum $k$ at three values of the Fermi momentum $k_F$.
\par\noindent
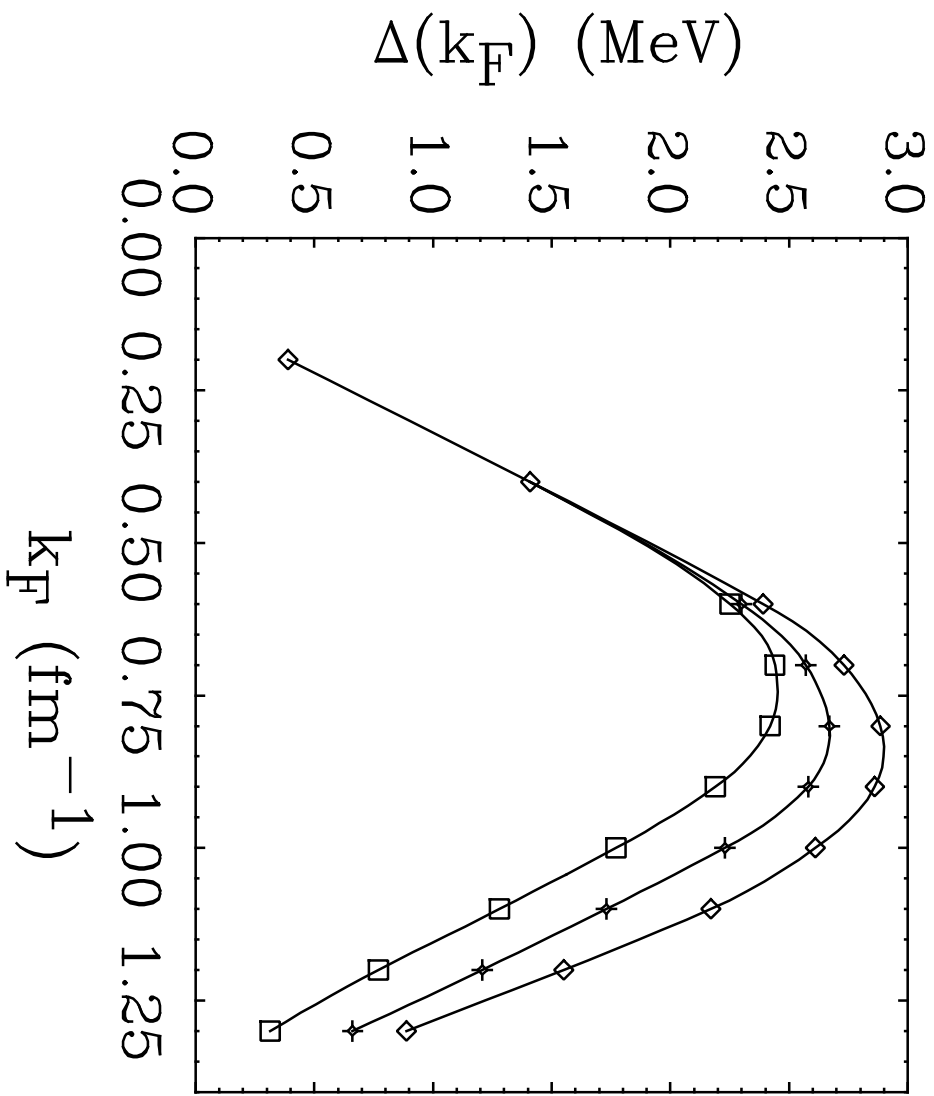
Fig. 4.- The superfluid gap value, at the Fermi momentum, as a function
of density, in the case of free single particle spectrum (diamonds),
with the inclusion of the factor $Z_k$ (crosses) and with the
inclusion of both $Z_k$ factor and the self-energy in the single particle 
spectrum (squares).


\newpage
\begin{thebibliography}{99}
\bibitem{shap} S.L. Shapiro, and S.A. Teukolsky, 
                 {\em Black Holes, White Dwarfs, and Neutron Stars} 
                 (John Wiley, New York, 1983).
\bibitem{pines} D. Pines and M.A. Alpar, Nature {\bf 316} (1985) 27.
\bibitem{baiko}  D.A. Baiko, and D.G. Yakolev, 
                 astro-ph/9812071, Astronomy and Astrophysics, in press. 
\bibitem{clark1}
  J. M. C. Chen, J. W. Clark, R. D. Dav\'e, and V. V. Khodel
  Nucl. Phys. {\bf A555} (1993) 59.
\bibitem{clark2} J. W. Clark, C.-G. K\"allman, C.-H. Yang, and D. A. Chakkalakal,
  Phys. Lett. {\bf 61B} (1976) 331.
\bibitem{hans} H.-J. Schulze, J. Cugnon, A. Lejeune, M. Baldo and U. Lombardo,
Phys. Lett. {\bf B375} (1996) 1. 
\bibitem{wambach}
  T. L. Ainsworth, J. Wambach, and D. Pines,
  Phys. Lett. {\bf B222} (1989)  173 ; \newline 
 J. Wambach, T. L. Ainsworth, and D. Pines,
  Nucl. Phys. {\bf A555} (1993)  128.
\bibitem{bab} S.-O. B\"ackman and G. E. Brown and J. A. Niskanen, Phys. Rep.
{\bf 124} (1985) 1.
\bibitem{baldo} M. Baldo, J. Cugnon, A. Lejeune and U. Lombardo,
 Nucl. Phys. {\bf A515} (1990) 409 ;  Nucl. Phys. {\bf A536} (1991) 349.
\bibitem{martino} M. Baldo,\O . Elgar\o y, L. Engvik, M. Hjorth-Jensen and 
H.J. Schulze, Phys. Rev. {\bf C58} (1998) 1921.  
\bibitem{bozek} P. Bozek, Nucl. Phys. {\bf 657} (1999) 187.                
\bibitem{BCS}  J. Bardeen, L.N. Cooper and J.R. Schrieffer, 
 Phys. Rev. {\bf108} (1957) 1175.
\bibitem{noz} P. Nozi\`eres, {\it Theory of Interacting Fermi Systems}, 
    W.A. Benjamin, New York, 1966.
\bibitem{abri} A. A. Abrikosov, L. P. Gor'kov~L~P and I. E. Dzyaloshinskii,
 {\it Methods of quantum field theory in statistical physics},
    (London: Prentice-Hall) 1963.
\bibitem{book} For a pedagogical introduction, see 
 {\it Nuclear Methods and the Nuclear Equation of State}, 
 Edited by M. Baldo, World Scientific, Singapore, International Review of
 Nuclear Physics Vol. 9, 1999.
\bibitem{fetter} A.L. Fetter and J.D. Walecka, {\it Quantum Theory of 
Many-Particle Systems}, Mc Graw-Hill, 1971.
\bibitem{neg} See e.g. J. W. Negele and H. Orland, 
 {\it Quantum Many-Particle Systems}, Frontiers in Physics Vol. 68,
 Addison Wisely 1987.
\bibitem{migdal} A.B. Migdal, {\it Theory of Finite Fermi Systems 
and Applications to Atomic Nuclei}, Interscience, London, 1967. 
\bibitem{old} M. Baldo, G. Giansiracusa, U. Lombardo and R. Pucci,
 Nuovo Cimento {\bf B58} (1980) 301.
\bibitem{wolfle} W\"olfle, Physica {\bf B90} (1977) 96.
\bibitem{private} P. Bozek, private communication.
\bibitem{mahaux} M. Baldo, G. Giansiracusa, U. Lombardo, C. Mahaux and
 R. Sartor, Nucl. Phys. {\bf A545} (1992) 741.


\end{thebibliography}
\end{document}